\documentclass[conference]{IEEEtran}

\usepackage{amsmath}
\usepackage{amssymb}
\usepackage[dvips]{graphicx}
\usepackage{setspace}
\usepackage{epsfig}
\newcommand{\K}{{\mathbf{K}}}
\newcommand{\x}{\mathbf{x}}
\newcommand{\y}{\mathbf{y}}

\newcommand{\n}{\mathbf{n}}
\newcommand{\hh}{\mathbf{H}}

\newcommand{\I}{\mathbf{I}}

\newcommand{\E}{\mathbb{E}}

\newcommand{\dd}{\dagger}

\newcommand{\tr}{{\text{tr\,}}}
\newcommand{\0}{\mathbf{0}}
\newcommand{\bu}{\mathbf{u}}
\newcommand{\U}{\mathbf{U}}
\newcommand{\bLambda}{\mathbf{\Lambda}}
\newcommand{\bPhi}{\mathbf{\Phi}}
\newcommand{\lm}{\lambda_{\max}}

\newcommand{\dc}{\dot{C}_s(0)}
\newcommand{\ddc}{\ddot{C}_s(0)}
\newcommand{\so}{\mathcal{S}_0}

\newcommand{\tsnr}{{\text{\footnotesize{SNR}}}}

\newtheorem{theo}{Theorem}
\newtheorem{prop}{Proposition}
\newtheorem{rem}{Remark}
\newtheorem{cor}{Corollary}

\begin{document}

\title{Secure Communication in the Low-SNR Regime: A Characterization of the Energy-Secrecy Tradeoff
}



%
\author{\authorblockN{Mustafa Cenk Gursoy}
\authorblockA{Department of Electrical Engineering\\
University of Nebraska-Lincoln, Lincoln, NE 68588\\ Email:
gursoy@engr.unl.edu}}


\maketitle

\begin{abstract}\footnote{This work was supported in part by the NSF
CAREER Grant CCF-0546384.} Secrecy capacity of a multiple-antenna wiretap channel is studied in the low signal-to-noise ratio ($\tsnr$) regime. Expressions for the first and second derivatives of the secrecy capacity with respect to $\tsnr$ at $\tsnr = 0$ are derived. Transmission strategies required to achieve these derivatives are identified. In particular, it is shown that it is optimal in the low-$\tsnr$ regime to transmit in the maximum-eigenvalue eigenspace of $\bPhi = \hh_m^\dd \hh_m - \frac{N_m}{N_e}\hh_e^\dd \hh_e$ where $\hh_m$ and $\hh_e$ denote the channel matrices associated with the legitimate receiver and eavesdropper, respectively, and $N_m$ and $N_e$ are the noise variances at the receiver and eavesdropper, respectively. Energy efficiency is analyzed by finding the minimum bit energy required for secure and reliable communications, and the wideband slope. Increased bit energy requirements under secrecy constraints are quantified. Finally, the impact of fading is investigated.
\end{abstract}

\section{Introduction} \label{sec:intro}

Secure transmission of confidential messages is a critical issue in communication systems and especially in wireless systems due to the broadcast nature of wireless transmissions. In \cite{Wyner}, Wyner addressed the transmission security from an information-theoretic point of view, and identified the rate-equivocation region and established the secrecy capacity of the discrete memoryless wiretap channel in which the wiretapper receives a degraded version of the signal observed by the legitimate receiver. The secrecy capacity is defined as the maximum communication rate from the transmitter to the legitimate receiver, which can be achieved while keeping the eavesdropper completely ignorant of the transmitted messages. Later, these results are extended to Gaussian wiretap channel in \cite{Hellman}. In \cite{Csiszar}, Csisz\'ar and K\"{o}rner considered a more general wiretap channel model and established the secrecy capacity when the transmitter has a common message for two receivers and a confidential message to only one. Recently, there has been a flurry of activity in the area of information-theoretic security, where, for instance, the impact of fading, cooperation, and interference on secrecy are studied (see e.g., \cite{specialissue} and the articles and references therein). Several recent results also addressed the secrecy capacity when multiple-antennas are employed by the transmitter, receiver, and the eavedropper \cite{Hero}--\cite{Oggier}. The secrecy capacity for the most general case in which arbitrary number of antennas are present at each terminal has been established in \cite{Khisti} and \cite{Oggier}.

In addition to security issues, another pivotal concern in most wireless systems is energy-efficient operation especially when wireless units are powered by batteries. From an information-theoretic perspective, energy efficiency can be
measured by the energy required to send one information bit
reliably. It is well-known that for unfaded and fading Gaussian channels subject to
average input power constraints, energy efficiency improves as one operates at lower $\tsnr$ levels, and the minimum bit energy is achieved as  $\tsnr$ vanishes \cite{Verdu}. Hence, requirements on energy efficiency necessitate operation in the low-$\tsnr$ regime. Additionally, operating at low $\tsnr$ levels has its benefits in terms of limiting the interference in wireless systems.

In this paper, in order to address the two critical issues of security and energy-efficiency jointly, we study the secrecy capacity in the low-$\tsnr$ regime. We consider a general multiple-input and multiple-output (MIMO) channel model and identify the optimal transmission strategies in this regime under secrecy constraints. Since secrecy capacity is in general smaller than the capacity attained in the absence of confidentiality concerns, energy per bit requirements increase due to secrecy constraints. In this work, we quantify these increased energy costs and address the energy-secrecy tradeoff.

\section{Channel Model} \label{sec:channelmodel}

We consider a MIMO channel model and assume that the transmitter, legitimate receiver, and eavesdropper are equipped with $n_T, n_R$, and $n_E$ antennas, respectively. We further assume that the channel input-output relations between the transmitter and legitimate receiver, and the transmitter and eavesdropper are given by
\begin{gather}
\y_m = \hh_m \x + \n_m \quad \text{ and } \quad 
\y_e = \hh_e \x + \n_e, \label{eq:model1}
\end{gather}
respectively. Above, $\x$ denotes the $n_T \times 1$--dimensional transmitted signal vector. This channel input is subject to the following average power constraint:
\begin{gather}
\E\{\|\x\|^2\} = \tr(\K_x) \le P
\end{gather}
where $\tr$ denotes the trace operation and $\K_x = E\{\x \x^\dd\}$ is the covariance matrix of the input. In (\ref{eq:model1}), $n_R \times 1$--dimensional $\y_m$ and $n_E \times 1$--dimensional $\y_e$ represent the received signal vectors at the legitimate receiver and eavesdropper, respectively. Moreover, $\n_m$ with dimension $n_R \times 1$ and $\n_e$ with dimension $n_E \times 1$ are independent, zero-mean Gaussian random vectors with $E\{\n_m \n_m^\dd\} = N_m \I$ and $E\{\n_e \n_e^\dd\} = N_e \I$, where $\I$ is the identity matrix. The signal-to-noise ratio is defined as
\begin{gather} \label{eq:snr}
\tsnr = \frac{\E\{\|\x\|^2\}}{\E\{\|\n_m\|^2\}} = \frac{P}{n_R N_m}.
\end{gather}
Finally, in the channel models, $\hh_m$ is the $n_R \times n_T$--dimensional channel matrix between the transmitter and legitimate receiver, and  $\hh_e$ is the $n_E \times n_T$--dimensional channel matrix between the transmitter and eavesdropper. While being fixed deterministic matrices in unfaded channels, $\hh_m$ and $\hh_e$ in fading channels are random matrices whose components denote the fading coefficients between the corresponding antennas at the transmitting and receiving ends.

\section{Secrecy in the Low-SNR Regime}

Recently, in \cite{Khisti} and \cite{Oggier}, it has been shown that when the channel matrices $\hh_m$ and $\hh_e$ are fixed for the entire transmission period and are known to all three terminals, then the secrecy capacity in nats per dimension is given by\footnote{Unless stated otherwise, \!all \!logarithms throughout the paper are to the base $e$.}
\begin{align}\label{eq:secrecycap}
C_s = &\frac{1}{n_R}\max_{\substack{\K_x \succeq \0 \\ \tr(\K_x) \le P}}  \log \det \left(\I + \frac{1}{N_m} \hh_m \K_x \hh_m^\dd\right)\nonumber
\\
&\hspace{2.2cm}-\log \det \left(\I + \frac{1}{N_e} \hh_e \K_x \hh_e^\dd\right)
\end{align}
where the maximization is over all possible input covariance matrices $\K_x \succeq \0$\footnote{$\succeq$ and $\succ$ denote positive semidefinite and positive definite partial orderings, respectively, for Hermitian matrices. If $\mathbf{A} \succeq \mathbf{B}$, then $\mathbf{A} - \mathbf{B}$ is a positive semidefinite matrix. Similarly, $\mathbf{A} \succ \mathbf{B}$ implies that $\mathbf{A} - \mathbf{B}$ is positive definite.} subject to a trace constraint. We note that since $\log \det \left(\I + 1/N_m \hh_m \K_x \hh_m^\dd \right)$ is a concave function of $\K_x$, the objective function in (\ref{eq:secrecycap}) is in general neither concave nor convex in $\K_x$, making the identification the optimal input covariance matrix a difficult task.

In this paper, we concentrate on the low-$\tsnr$ regime. In this regime, the behavior of the secrecy capacity can be accurately predicted by its first and second derivatives with respect to $\tsnr$ at $\tsnr = 0$:
\begin{align}
C_s(\tsnr) &= \dot{C}_s(0) \tsnr + \frac{\ddot{C}_s(0)}{2} \tsnr^2 + o(\tsnr^2). \label{eq:lowsnrapprox}
\end{align}
Moreover, $\dot{C}_s(0)$ and $\ddot{C}_s(0)$ also enable us to analyze the energy efficiency in the low-$\tsnr$ regime through \cite{Verdu}
\begin{gather}\label{eq:ebno-so}
\frac{E_b}{N_0}_{s,\min} = \frac{\log 2}{\dc} \text{ and } \so = \frac{2 \left[\dc\right]^2}{-\ddc}
\end{gather}
where $\frac{E_b}{N_0}_{s,\min}$ denotes the minimum bit energy required for reliable communication under secrecy constraints, and $\so$ denotes the wideband slope
which is the slope of the secrecy capacity in bits/dimension/(3 dB) at the point $\frac{E_b}{N_0}_{s,\min}$.
These quantities provide a linear
approximation of
the secrecy capacity in the low-$\tsnr$ regime. 
While $\frac{E_b}{N_0}_{s,\min}$ is a performance measure for vanishing $\tsnr$, $\so$ together with $\frac{E_b}{N_0}_{s,\min}$ characterize the performance at low but nonzero $\tsnr$s. We note that the formula for the minimum bit energy is valid if $C_s$ is a concave function of $\tsnr$, which we show later in the paper.

The following result identifies the first and second derivatives of the secrecy capacity at $\tsnr = 0$.

\begin{theo} \label{theo:secrecyderivatives}
The first derivative of the secrecy capacity in (\ref{eq:secrecycap}) with respect to $\tsnr$ at $\tsnr = 0$ is
\begin{gather}\label{eq:capfirstderiv}
\dot{C}_s(0) = [\lm(\bPhi)]^+ = \left\{
\begin{array}{ll}
\lm(\bPhi) & \text{if } \lm(\bPhi)>0
\\
0 & \text{else}
\end{array}\right.
\end{gather}
where
$
\bPhi = \hh_m^\dd \hh_m - \frac{N_m}{N_e}\hh_e^\dd \hh_e.
$
Moreover, the second derivative of the secrecy capacity at $\tsnr = 0$ is given by
\begin{align}
\ddot{C}_s(0) = -n_R \min_{\substack{\{\alpha_i\} \\ \alpha_i \in [0,1] \, \forall i \\ \sum_{i=1}^l \alpha_i = 1}} \sum_{i,j = 1}^l &\alpha_i \alpha_j \bigg( |\bu_j^\dd \hh_m^\dd \hh_m \bu_i|^2
\nonumber
\\ 
&\hspace{-1.75cm}- \frac{N_m^2}{N_e^2} |\bu_j^\dd \hh_e^\dd \hh_e \bu_i|^2\bigg) 1\{\lm(\bPhi > 0)\} \label{eq:capsecondderiv}
\end{align}
where $l$ is the multiplicity of $\lm(\bPhi)>0$, $\{\bu_i\}$ are the eigenvectors that span the maximum-eigenvalue eigenspace, and $1\{\lm(\bPhi) > 0 \} = \left\{
\begin{array}{ll}
1 & \text{if } \lm(\bPhi) > 0
\\
0 & \text{else}
\end{array}\right.
$ is the indicator function.
\end{theo}

\emph{Proof}: We first note that the input covariance matrix $\K_x = E\{\x \x^\dd\}$ is by definition a positive semidefinite Hermitian matrix. As a Hermitian matrix, $\K_x$ can be written as \cite[Theorem 4.1.5]{matrixbook}
\begin{align} \label{eq:spectral}
\K_x = \U \bLambda \U^\dd
\end{align}
where $\U$ is a unitary matrix and $\bLambda$ is a real diagonal matrix. Using (\ref{eq:spectral}), we can also express $\K_x$ as
\begin{gather}
\K_x = \sum_{i = 1}^{n_T} d_i \bu_i \bu_i^\dd
\end{gather}
where $\{d_i\}$ are the diagonal components of $\bLambda$, and $\{\bu_i\}$ are the column vectors of $\U$ and form an orthonormal set. Assuming that the input uses all the available power, we have $\tr(\K_x) = \sum_{i = 1}^{n_T}{d_i} = P$. Noting that $\K_x$ is positive semidefinite and hence  $d_i \ge 0$, we can write $d_i = \alpha_i P$ where $\alpha_i \in [0,1] $ $\forall i$ and $\sum_{i = 1}^{n_T} \alpha_i = 1$. Now, the secrecy rate achieved with a particular covariance matrix $\K_x$ can be expressed as
\begin{align}
I_s(\tsnr) &= \frac{1}{n_R} \Bigg( \log \det \left(\I + n_R \, \tsnr \sum_{i = 1}^{n_T} \alpha_i \hh_m \bu_i \bu_i^\dd \hh_m^\dd\right) \nonumber
\\
&-\log \det \left(\I + \frac{n_R N_m}{N_e} \, \tsnr \sum_{i = 1}^{n_T} \alpha_i \hh_e \bu_i \bu_i^\dd \hh_e^\dd\right) \Bigg).
\end{align}
where $\tsnr$ is defined in (\ref{eq:snr}). As also noted in \cite{Verdu}, we can easily show that
\vspace{-.5cm}
\begin{align}
\frac{d}{dv} \log \det (\I + v \mathbf{A}) |_{v = 0} &= \tr(\mathbf{A}), \label{eq:derivlogdet}
\\
\frac{d^2}{dv^2} \log \det (\I + v \mathbf{A}) |_{v = 0} &= - \tr(\mathbf{A}^2). \label{eq:deriv2logdet}
\end{align}
Now, using (\ref{eq:derivlogdet}), we obtain the following expression for the first derivative of the secrecy rate $I_s$ with respect to $\tsnr$ at $\tsnr = 0$:
\begin{align}
\hspace{-.2cm}\dot{I}_s(0) &= \sum_{i = 1}^{n_T} \alpha_i \left(\tr(\hh_m \bu_i \bu_i^\dd \hh_m^\dd) - \frac{N_m}{N_e}\tr(\hh_e \bu_i \bu_i^\dd \hh_e^\dd) \right)
\\
&= \sum_{i = 1}^{n_T} \alpha_i \left(\bu_i^\dd \hh_m^\dd \hh_m \bu_i  - \frac{N_m}{N_e}\bu_i^\dd \hh_e^\dd \hh_e \bu_i  \right) \label{eq:dotI2}
\\
&= \sum_{i = 1}^{n_T} \alpha_i \bu_i^\dd \left(\hh_m^\dd \hh_m - \frac{N_m}{N_e}\hh_e^\dd \hh_e\right) \bu_i
= \sum_{i = 1}^{n_T} \alpha_i \bu_i^\dd  \bPhi \bu_i \label{eq:dotI4}
\end{align}
where (\ref{eq:dotI2}) follows from the property that $\tr(\mathbf{A}\mathbf{B}) = \tr(\mathbf{B}\mathbf{A})$. Also, in (\ref{eq:dotI4}), we have defined $\bPhi = \hh_m^\dd \hh_m - \frac{N_m}{N_e}\hh_e^\dd \hh_e$. Since $\bPhi$ is a Hermitian matrix and $\{\bu_i\}$ are unit vectors, we have \cite[Theorem 4.2.2]{matrixbook}
\begin{gather} \label{eq:upperbound-lambda}
\bu_i^\dd  \bPhi \bu_i \le \lambda_{\max}(\bPhi) \quad \forall i
\end{gather}
where $\lambda_{\max}(\bPhi)$ denotes the maximum eigenvalue of the matrix $\bPhi$. Recall that $\alpha_i \in [0,1]$ and $\sum_{i} \alpha_i = 1$. Then, from (\ref{eq:upperbound-lambda}), we obtain
\vspace{-.3cm}
\begin{align}
\dot{I}_s(0) = \sum_{i = 1}^{n_T} \alpha_i \bu_i^\dd  \bPhi \bu_i \le \lambda_{\max}(\bPhi). \label{eq:upperbounddotI}
\end{align}
Note that this upper bound can be achieved if, for instance, $\alpha_1 = 1$ and $\alpha_i = 0$ $\forall i \neq 1$, and $\bu_1$ is chosen as the eigenvector that corresponds to the maximum eigenvalue of $\bPhi$. Heretofore, we have implicitly assumed that $\lm(\bPhi) > 0$ and all the available power is used to transmit the information in the direction of the maximum eigenvalue. If $\lm(\bPhi) \le 0$, then all eigenvalues of $\bPhi$ are less than or equal to zero, and hence $\bPhi$ is a negative semidefinite matrix. In this situation, none of the channels of the legitimate receiver is stronger than those corresponding ones of the eavesdropper. In such a case, secrecy capacity is zero. Therefore, if $\lm(\bPhi) \le 0$, we have $\dot{C}_s(0) = 0$. Finally, we conclude from (\ref{eq:upperbounddotI}) and the above discussion that the first derivative of the secrecy capacity with respect to $\tsnr$ at $\tsnr = 0$ is given by
\begin{align}
\dot{C}_s(0) = [\lm(\bPhi)]^+ = \left\{
\begin{array}{ll}
\lm(\bPhi) & \text{if } \lm(\bPhi)>0
\\
0 & \text{else}
\end{array}\right..
\end{align}
If $\lm(\bPhi) > 0$ is distinct, $\dot{C}_s(0)$ is achieved when we choose $\K_x = P \bu_1 \bu_1^\dd$ where $\bu_1$ is the eigenvector that corresponds to $\lm(\bPhi)$. Therefore, beamforming in the direction in which the eigenvalue of $\bPhi$ is maximized is optimal in the sense of achieving the first derivative of the secrecy capacity in the low-$\tsnr$ regime. More generally, if $\lm(\bPhi) > 0 $ has a multiplicity, any covariance matrix in the following form achieves the first derivative: \vspace{-.3cm}
\begin{gather} \label{eq:covariance-opt}
\K_x = P \sum_{i = 1}^{l} \alpha_i \bu_i \bu_i^\dd
\end{gather}
where $l$ is the multiplicity of the maximum eigenvalue, $\{\bu_i\}_{i = 1}^l$ are the eigenvectors that span the maximum-eigenvalue eigenspace, and $\{\alpha_i\}_{i = 1}^l$ are constants, taking values in $[0,1]$ and having the sum $\sum_{i=1}^l \alpha_i = 1$. Therefore, transmission in the maximum-eigenvalue eigenspace is necessary to achieve $\dot{C}_s(0)$.

Next, we consider the second derivative of the secrecy capacity. Again, when $\lm(\bPhi) \le 0$, the secrecy capacity is zero and therefore $\ddc = 0$. Hence, in the following, we consider the case in which $\lm(\bPhi) > 0$. Suppose that the input covariance matrix is chosen as in (\ref{eq:covariance-opt}) with a particular set of $\{\alpha_i\}$. Then, using (\ref{eq:deriv2logdet}), we can obtain
\begin{align}
\ddot{I_s}(0) &= -n_R \,\, \tr \left( \left( \sum_{i = 1}^{l} \alpha_i \hh_m \bu_i  \bu_i^\dd \hh_m^\dd \right)^2\right) \nonumber
\\
& \hspace{0.5cm}+ n_R \frac{N_m^2}{N_e^2} \,\, \tr\left( \left( \sum_{i = 1}^{l} \alpha_i \hh_e \bu_i  \bu_i^\dd \hh_e^\dd \right)^2\right)
\\
&= -n_R \sum_{i,j} \alpha_i \alpha_j \left( |\bu_j^\dd \hh_m^\dd \hh_m \bu_i|^2 - \frac{N_m^2}{N_e^2} |\bu_j^\dd \hh_e^\dd \hh_e \bu_i|^2\right) \label{eq:ratesecondderiv}
\end{align}
where  (\ref{eq:ratesecondderiv}) is obtained by using the fact that $\tr(\mathbf{A}\mathbf{B}) = \tr(\mathbf{B}\mathbf{A})$ and performing some straightforward manipulations. Note again that $\{\bu_i\}$ are the eigenvectors spanning the maximum-eigenvalue eigenspace of $\bPhi$. Being necessary to achieve the first derivative, the covariance structure given in (\ref{eq:covariance-opt}) is also necessary to achieve the second derivative. Therefore, the second derivative of the secrecy capacity at $\tsnr=0$ is the maximum of the expression in (\ref{eq:ratesecondderiv}) over all possible values of $\{\alpha_i\}$. Hence,
\begin{align}
\ddot{C}_s(0) = -n_R \!\!\!\!\min_{\substack{\{\alpha_i\} \\ \alpha_i \in [0,1] \, \forall i \\ \sum_{i=1}^l \alpha_i = 1}} \sum_{i,j} &\alpha_i \alpha_j \bigg( |\bu_j^\dd \hh_m^\dd \hh_m \bu_i|^2 - \frac{N_m^2}{N_e^2} |\bu_j^\dd \hh_e^\dd \hh_e  \bu_i|^2\bigg) \label{eq:capsecondderivintheproof}
\end{align}
Since $\ddc$ is equal to the expression in (\ref{eq:capsecondderivintheproof}) when $\lm(\bPhi) > 0$ and is zero otherwise,  the final expression in (\ref{eq:capsecondderiv}) is obtained by multiplying the formula in (\ref{eq:capsecondderivintheproof}) with the indicator function $1\{\lm(\bPhi) > 0\}$. \hfill $\blacksquare$
\begin{rem} \label{rem:nosecrecy}
 In the absence of secrecy constraints, the first and second derivatives of the MIMO capacity at $\tsnr = 0$ are \cite{Verdu}
\begin{align}
\dot{C}(0) = \lm(\hh_m^\dd \hh_m) \text{ and }
\ddot{C}(0) = -\frac{n_R}{l}\lm^2(\hh_m^\dd \hh_m)
\end{align}
where $l$ is the multiplicity of $\lm(\hh_m^\dd \hh_m)$. Hence, the first and second derivatives are achieved by transmitting in the maximum-eigenvalue eigenspace of $\hh_m^\dd \hh_m$, the subspace in which the transmitter-receiver channel is the strongest. Due to the optimality of the water-filling power allocation method, power should be equally distributed in each orthogonal direction in this subspace in order for the second derivative to be achieved.
\end{rem}

\begin{rem}
We see from Theorem \ref{theo:secrecyderivatives} that when there are secrecy constraints, we should at low $\tsnr$s transmit in the direction in which the transmitter-receiver channel is strongest \emph{with respect to the transmitter-eavesdropper channel} normalized by the ratio of the noise variances. For instance, $\dot{C}_s(0)$ can be achieved by beamforming in the direction in which the eigenvalue of $\bPhi$ is maximized. On the other hand, if $\lm(\bPhi)$ has a multiplicity, the optimization problem in (\ref{eq:capsecondderiv}) should be solved to identify how the power should be allocated to different orthogonal directions in the maximum-eigenvalue eigenspace so that the second-derivative $\ddot{C}_s(0)$ is attained. In general, the optimal power allocation strategy is neither water-filling nor beamforming. For instance, consider parallel Gaussian channels for both transmitter-receiver and transmitter-eavesdropper links, and assume that $\hh_m^\dd \hh_m = \text{diag}(5, 4, 2)$ and $\hh_e^\dd \hh_e = \text{diag}(2, 1, 1)$ where $\text{diag()}$ is used to denote a diagonal matrix with components provided in between the parentheses. Assume further that the noise variances are equal, i.e., $N_m = N_e$. Then, it can be easily seen that $\lm(\bPhi) = 3$ and has a multiplicity of $2$. Solving the optimization problem in (\ref{eq:capsecondderiv}) provides $\alpha_1 = 5/12$ and $\alpha_2 = 7/12$. Hence, approximately, $42\%$ of the power is allocated to the channel for which the transmitter-receiver link has a strength of $5$, and $58\%$ is allocated for the channel with strength $4$.
\end{rem}

\begin{rem}
When $\lm(\bPhi) > 0$ is distinct, then beamforming in the direction in which $\lambda(\bPhi)$ is maximized is optimal in the sense of achieving both $\dot{C}_s(0)$ and $\ddot{C}_s(0)$. Moreover, in this case, we have
\begin{gather}
\ddot{C}_s(0)  = -n_R \left( \|\hh_m \bu_1\|^4 - \frac{N_m^2}{N_e^2} \|\hh_e \bu_1\|^4\right)
\end{gather}
where $\bu_1$ is the eigenvector that corresponds to $\lm(\bPhi)$.
\end{rem}

\begin{rem} \label{rem:eigenvaluediff}
From \cite[Theorem 4.3.1]{matrixbook}, we know that for two Hermitian matrices $\mathbf{A}$ and $\mathbf{B}$ with the same dimensions, we have
\begin{gather}
\lm(\mathbf{A}+\mathbf{B}) \le \lm(\mathbf{A}) + \lm(\mathbf{B}).
\end{gather}
Applying this result to our setting yields
\begin{gather}
\lm(\bPhi) \le \lm(\hh_m^\dd \hh_m) -\lambda_{\min}\left(\frac{N_m}{N_e}\hh_e^\dd \hh_e \right).
\end{gather}
Therefore, we conclude from Remark \ref{rem:nosecrecy} that secrecy constraints diminish the first derivative $\dot{C}_s(0)$ at least by a factor of $\lambda_{\min}\left(\frac{N_m}{N_e}\hh_e^\dd \hh_e \right)$ when compared to the case in which there are no such constraints.
\end{rem}

\begin{rem}
In the case in which each terminal has a single antenna, the results of Theorem \ref{theo:secrecyderivatives} specialize to
\begin{align}
\dot{C}_s(0) &= \left[|h_m|^2 - \frac{N_m}{N_e}|h_e|^2\right]^+
\\
\ddot{C}_s(0) &= -\left[ |h_m|^4 - \frac{N_m^2}{N_e^2} |h_e|^4\right]^+.
\end{align}
\end{rem}

In the next result, we show that the secrecy capacity is concave in $\tsnr$.

\begin{prop} \label{prop:concavity}
The secrecy capacity $C_s$ achieved under the average power constraint $\E\{\|\x\|^2\} \le P$ is a concave function of $\tsnr$.
\end{prop}

\emph{Proof:} Concavity can be easily shown using the time-sharing argument. Assume that at power level $P_1$ and signal-to-noise ratio $\tsnr_1$, the optimal input is $\x_1$, which satisfies $\E\{\|\x_1\|^2\} \le P_1$, and the secrecy capacity is $C_s(\tsnr_1)$. Similarly, for $P_2$ and $\tsnr_2$, the optimal input is $\x_2$, which satisfies $\E\{\|\x_2\|^2\} \le P_2$, and the secrecy capacity is $C_s(\tsnr_2)$. Now, we assume that the transmitter performs time-sharing by transmitting at two different power levels using $\x_1$ and $\x_2$. More specifically, in $\theta$ fraction of the time, the transmitter uses the input $\x_1$, transmits at most at $P_1$, and achieves the secrecy rate $C_s(\tsnr_1)$. In the remaining $(1-\theta)$ fraction of the time, the transmitter employs $\x_2$, transmits at most at $P_2$, and achieves the secrecy rate $C_s(\tsnr_2)$. Hence, this scheme overall achieves the average secrecy rate of
\begin{gather} \label{eq:timesharingrate}
\theta C_s(\tsnr_1) + (1-\theta) C_s(\tsnr_2)
\end{gather}
by transmitting at the level $\theta \E\{\|\x_1\|^2\} + (1-\theta) \E\{\|\x_2\|^2\}  \le P_\theta = \theta P_1 + (1-\theta) P_2$.  The average signal-to-noise ratio is $\tsnr_\theta = \theta \tsnr_1 + (1-\theta) \tsnr_2$. Therefore, the secrecy rate in (\ref{eq:timesharingrate}) is an achievable secrecy rate at $\tsnr_\theta$. Since the secrecy capacity is the maximum achievable secrecy rate, the secrecy capacity at $\tsnr_\theta$ is larger than that in (\ref{eq:timesharingrate}), i.e.,
\begin{align}
C_s(\tsnr_\theta) &= C_s(\theta \tsnr_1 + (1-\theta) \tsnr_2)
\\
&\ge \theta C_s(\tsnr_1) + (1-\theta) C_s(\tsnr_2),
\end{align}
showing the concavity. \hfill $\blacksquare$

We further note that the concavity can also be shown using the following facts. As also discussed in \cite{Liu-Shamai}, MIMO secrecy capacity is obtained by proving in the converse argument that the considered upper bound is tight and
\begin{gather}
C_s = \max_{p(\x)} \min_{p(\y_r^{'}, \y_e^{'} | \x) \in \mathcal{D} } I(\x ; \y_r^{'} | \y_e^{'})
\end{gather}
where $\mathcal{D}$ is the set of joint conditional density functions $p(\y_r^{'}, \y_e^{'} | \x)$ that satisfy $p(\y_r^{'}| \x) = p(\y_r | \x)$ and $p(\y_e^{'}| \x) = p(\y_e | \x)$. Note that for fixed channel distributions, the mutual information $I(\x ; \y_r^{'} | \y_e^{'})$ is a concave function of the input distribution $p(\x)$. Since the pointwise infimum of a set of concave functions is concave \cite{convex}, $f(p(\x)) = \min_{p(\y_r^{'}, \y_e^{'} | \x) \in \mathcal{D} } I(\x ; \y_r^{'} | \y_e^{'})$ is also a concave function of $p(\x)$. Concavity of the functional $f$ and the fact that maximization is over input distributions satisfying $\E\{\|\x\|^2\} \le P$ lead to the concavity of the secrecy capacity with respect to $\tsnr$.

We can now write the following corollary to Proposition \ref{prop:concavity} and Theorem \ref{theo:secrecyderivatives}.

\begin{cor}
The minimum bit energy attained under secrecy constraints is
\begin{gather}
\frac{E_b}{N_0}_{s,\min} =  \frac{\log 2} {[\lm(\bPhi)]^+}.
\end{gather}
\end{cor}

\begin{rem}
From Remark \ref{rem:eigenvaluediff}, we can write
\begin{align}
\frac{E_b}{N_0}_{s,\min} = \frac{\log 2} {[\lm(\bPhi)]^+} &\ge \frac{\log 2} {\lm(\hh_m^\dd \hh_m) -\lambda_{\min}\left(\frac{N_m}{N_e}\hh_e^\dd \hh_e \right)}\nonumber
\\
&\ge \frac{\log 2} {\lm(\hh_m^\dd \hh_m)} = \frac{E_b}{N_0}_{\min} \label{eq:ebnomin_nosecrecy}
\end{align}
where $\frac{E_b}{N_0}_{\min}$ in (\ref{eq:ebnomin_nosecrecy}) denotes the minimum bit energy in the absence of secrecy constraints. Hence, in general, secrecy requirements increase the energy expenditure. When secure communication is not possible, $[\lm(\bPhi)]^+ = 0$ and $\frac{E_b}{N_0}_{s,\min} = \infty$.

The expression for the wideband slope $S_0$ can be readily obtained by plugging in the expressions in (\ref{eq:capfirstderiv}) and (\ref{eq:capsecondderiv}) into that in (\ref{eq:ebno-so}).

\end{rem}

\begin{rem}
Energy costs of secrecy can easily be identified in the single-antenna case. Clearly, the minimum bit energy in the presence of secrecy is strictly greater than that in the absence of such constraints:
\begin{gather}
\frac{E_b}{N_0}_{s,\min} = \frac{\log 2}{\left[|h_m|^2 - \frac{N_m}{N_e} |h_e|^2\right]^+} > \frac{\log 2}{|h_m|^2} = \frac{E_b}{N_0}_{\min}
\end{gather}
when  $\frac{N_m}{N_e} |h_e|^2 > 0 $. Furthermore, the energy requirement increases monotonically as the value of $\frac{N_m}{N_e} |h_e|^2$ increases. Indeed, when $\frac{N_m}{N_e} |h_e|^2 = |h_m|^2$, secure communication is not possible and $\frac{E_b}{N_0}_{s,\min} = \infty$.
\end{rem}

\section{The Impact of Fading}

In this section, we assume that the channel matrices $\hh_m$ and $\hh_e$ are random matrices whose components are ergodic random variables, modeling fading in wireless transmissions. We again assume that realizations of these matrices are perfectly known by all the terminals. As discussed in \cite{Liang}, fading channel can be regarded as a set of parallel subchannels each of which corresponds to a particular fading realization. Hence, in each subchannel, the channel matrices are fixed similarly as in the channel model considered in the previous section. In \cite{Liang}, Liang \emph{et al.} have shown that having independent inputs for each subchannel is optimal and the secrecy capacity of the set of parallel subchannels is equal to the sum of the capacities of subchannels. Therefore, the secrecy capacity of fading channels can be be found by averaging the secrecy capacities attained for different fading realizations.

We assume that the transmitter is subject to a short-term power constraint. Hence, for each channel realization, the same amount of power is used and we have $\tr(\K_x) \le P$. With this assumption, the transmitter is allowed to perform power adaptation in space across the antennas, but not across time. Under such constraints, it can easily be seen from the above discussion that the average secrecy capacity in fading channels is given by
\begin{align}\label{eq:avgsecrecycap}
C_s = &\frac{1}{n_R} \, \E_{\hh_m, \hh_e} \Bigg\{\max_{\substack{\K_x \succeq \0 \\ \tr(\K_x) \le P}}  \log \det \left(\I + \frac{1}{N_m} \hh_m \K_x \hh_m^\dd\right)\nonumber
\\
&\hspace{2.6cm}-\log \det \left(\I + \frac{1}{N_e} \hh_e \K_x \hh_e^\dd\right)\Bigg\}
\end{align}
where the expectation is with respect to the joint distribution of $(\hh_m, \hh_e)$. Note that the only difference between (\ref{eq:secrecycap}) and (\ref{eq:avgsecrecycap}) is the presence of expectation in (\ref{eq:avgsecrecycap}). Due to this similarity, the following result can be obtained immediately as a corollary to Theorem \ref{theo:secrecyderivatives}.
\begin{cor}
The first derivative of the average secrecy capacity in (\ref{eq:avgsecrecycap}) with respect to $\tsnr$ at $\tsnr = 0$ is
\begin{gather}
\dot{C}_s(0) = \E_{\hh_m, \hh_e}\{[\lm(\bPhi)]^+\}
\end{gather}
where again
$
\bPhi = \hh_m^\dd \hh_m - \frac{N_m}{N_e}\hh_e^\dd \hh_e.
$
The second derivative of the average secrecy capacity at $\tsnr = 0$ is given by
\begin{align} \label{eq:avgcapsecondderiv}
\ddot{C}_s(0) = -n_R \E_{\hh_m, \hh_e} &\Bigg\{\min_{\substack{\{\alpha_i\} \\ \alpha_i \in [0,1] \, \forall i \\ \sum_{i=1}^l \alpha_i = 1}} \!\!\sum_{i,j = 1}^l \alpha_i \alpha_j \bigg( |\bu_j^\dd \hh_m^\dd \hh_m \bu_i|^2  \nonumber
\\
& \hspace{-1.2cm}- \frac{N_m^2}{N_e^2} |\bu_j^\dd \hh_e^\dd \hh_e \bu_i|^2\bigg)  1\{\lm(\bPhi) > 0\} \Bigg\}
\end{align}
where $1\{\cdot\}$ again denotes the indicator function, $l$ is the multiplicity of $\lm(\bPhi)>0$, and $\{\bu_i\}$ are the eigenvectors that span the maximum-eigenvalue eigenspace for particular realizations of $\hh_m$ and $\hh_e$.
\end{cor}

\begin{rem}
Similarly as in the unfaded case, $\dot{C}_s(0)$ is achieved by always transmitting in the maximum-eigenvalue eigenspace of the realizations of the channel matrices $\hh_m$ and $\hh_e$. In order to achieve the second derivative, optimal values of $\{\alpha_i\}$ (or equivalently the optimal power allocation across the antennas) should be identified again for each possible realization of the channel matrices.
\end{rem}

\begin{rem}
In the single-antenna case in which $n_T = n_R = n_E = 1$, the first and second derivatives of the average secrecy capacity become 
\begin{align}
\dot{C}_s(0) &= \E_{h_m, h_e}\left\{\left[|h_m|^2 - \frac{N_m}{N_e}|h_e|^2\right]^+\right\}
\\
\ddot{C}_s(0) &= \E_{h_m, h_e}\left\{\left[|h_m|^4 - \frac{N_m}{N_e}|h_e|^4\right]^+\right\}.
\end{align}
\end{rem}

\begin{cor}
The minimum bit energy achieved in fading channels under secrecy constraints is
\begin{gather}
\frac{E_b}{N_0}_{s,\min} =  \frac{\log 2} {\E_{\hh_m, \hh_e}\{[\lm(\bPhi)]^+\}}.
\end{gather}
\end{cor}

\begin{rem}
Fading has a potential to improve the low-$\tsnr$ performance and hence the energy efficiency. To illustrate this, we consider the following example. Consider first the unfaded Gaussian channel in which the deterministic channel coefficients are $h_m = h_e = 1$. For this case, we have
\begin{align}
\dc = \left[1 - \frac{N_m}{N_e}\right]^+ \text{ and } \frac{E_b}{N_0}_{s,\min} = \frac{\log 2}{\left[1 - \frac{N_m}{N_e}\right]^+}.
\end{align}
Now, consider a Rayleigh fading environment and assume that $h_m$ and $h_e$ are independent, zero-mean, Gaussian random variables with variances $E\{|h_m|^2\} = E\{|h_e|^2\} = 1$. Then, we can easily find that
\begin{gather}
\dot{C}_s(0) = \E_{h_m, h_e}\left\{\left[|h_m|^2 - \frac{N_m}{N_e}|h_e|^2\right]^+\right\} = \frac{N_e}{N_m + N_e}
\end{gather}
leading to $\frac{E_b}{N_0}_{s,\min} = \frac{\log 2}{\frac{N_e}{N_m + N_e}}$.
Note that if $N_e > 0$,
$
\frac{N_e}{N_m + N_e} > \left[1 - \frac{N_m}{N_e}\right]^+.
$ Hence, fading strictly improves the low-$\tsnr$ performance by increasing $\dc$ and decreasing the minimum bit energy even without performing power control over time. Further gains are possible with power adaptation. Another interesting observation is the following. In unfaded channels, if $N_m \ge N_e$, the minimum bit energy is infinite and secure communication is not possible. On the other hand, in fading channels, the bit energy is finite as long as $N_m$ is finite and $N_e > 0$. Clearly, even if $N_m \ge N_e$, favorable fading conditions enable secure transmission in fading channels.

\end{rem}

%



\vspace{-.19cm}
\begin{thebibliography}{}


\bibitem{Wyner} A. D. Wyner, ``The wire-tap channel," \emph{Bell Syst. Tech. J.}, vol. 54, pp. 1355-1367, Oct. 1975

\bibitem{Hellman} S. K. Leung-Yan-Cheong and M. E. Hellman, ``The Gaussian wire-tap channel," \emph{IEEE
Trans. Inform. Theory}, vol.~24, pp.~451-456, Jul. 1978.

\bibitem{Csiszar} I. Csisz\'ar and J. K\"{o}rner , ``Broadcast channels with confidential messages," \emph{IEEE
Trans. Inform. Theory}, vol.~3, pp.~339-348, May 1978.

 \bibitem{specialissue} Special issue on information-theoretic security, \emph{IEEE Trans. Inform. Theory}, vol.~54, no. 6, June 2008.

\bibitem{Hero} A. O. Hero, ``Secure space-time communication," \emph{IEEE
Trans. Inform. Theory}, vol.~49, pp.~3235-3249, Dec. 2003.

\bibitem{Li} Z. Li, W. Trappe, and R. D. Yates, ``Secret communication via multi-antenna transmission," 41st Conference on Information Sciences and Systems (CISS), Baltimore, March 2007.

\bibitem{Shafiee} S. Shafiee, N. Liu, and S. Ulukus, ``Towards the secrecy capacity of the Gaussian MIMO wire-tap channel: The 2-2-1 Channel," submitted for publication. Also available at http://arxiv.org/abs/0709.3541.



\bibitem{Khisti} A. Khisti and G. W. Wornell, ``The MIMOME channel," Proc. of the 45th Annual Allerton Conference on Communication, Control, and Computing, October 2007. Also available at http://arxiv.org/abs/0710.1325.

\bibitem{Oggier} F. Oggier and B. Hassibi, ``The secrecy capacity of the MIMO Wiretap channel," available at http://arxiv.org/abs/0710.1920.

\bibitem{Liu-Shamai} T. Liu and S. Shamai (Shitz), ``A note on the secrecy capacity of the multi-antenna wiretap channel," submitted for publication. Also available at http://arxiv.org/abs/0710.4105.

\bibitem{Verdu} S. Verd\'u, ``Spectral efficiency in the wideband regime," \emph{IEEE
Trans. Inform. Theory}, vol.~48, pp.~1319-1343, June 2002.

\bibitem{Liang}  Y. Liang, H. V. Poor, and S. Shamai (Shitz),``Secure communication over fading channels," \emph{IEEE Trans. Inform. Theory}, vol.~54, pp.~2470 - 2492, June 2008.

%

\bibitem{matrixbook} R. A. Horn and C. R. Johnson, \emph{Matrix Analysis}, Cambridge University Press, 1999.

\bibitem{convex} S. Boyd and L. Vandenberghe, \emph{Convex
Optimization}, Cambridge University Press, 2004.

\end{thebibliography}
\end{document}